# AN EXAMINATION OF 'ATRI'S ECLIPSE' AS DESCRIBED IN THE RIG VEDA

**Mayank Vahia**
*Tata Institute of Fundamental Research (Retired), Mumbai 400 005, India.*
E-mail: vahia@tifr.res.in

**and**

**Mitsuru Sôma**
*National Astronomical Observatory of Japan, Mitaka, Tokyo, Japan.*
E-mail: Mitsuru.Soma@gmail.com

**Abstract:** The earliest written reference in Indian astronomy to a total solar eclipse is in the *Rig Veda* where Rishi Atri is said to have demolished the asura Swarbhanu to liberate the Sun from a total solar eclipse. The *Rig Veda* describes the occurrence of the eclipse, how the Sun suddenly disappeared in the daytime under the spell of the Asura. The people and gods were scared but the Great Sage Atri saved the Sun and restored his full glory. While discussing the eclipse, Tilak refers to the eclipse as having occurred when the Vernal Equinox was in Orion, and three days prior to the Autumnal Equinox. Based on these data, we identify 'Atri's eclipse' as the one that occurred on 22 October 4202 BC or on 19 October 3811 BC.

## 1 INTRODUCTION

The *Rig Veda* is one of the oldest known documents. It dates from 1500 BC, when its contents were assimilated and formalised on the basis of traditions of different schools of thought. It was essentially a summary of various religious ideas and philosophies, as well as their image of the world and its working as understood at that time. It comprises a set of 10 books associated with 10 different groups of priests who assimilated different aspects of the prevailing belief systems (see Dalal, 2014; Donigar, 1984). The writing style in the *Rig Veda* is highly poetic and abstract, and sometimes it is difficult to understand. It also requires some experience in order to interpret it.

While the *Rig Veda* dates from 1500 BC, there is a significant amount of evidence that it incorporates memories of events that were much further back in time. For example, it discusses events when the Vernal Equinox was in Orion, which occurred around 4500 BC, while the final reference to the Vernal Equinox in the *Rig Veda* relates to its being in the Pleiades, which happened in 2230 BC.

There are various other astronomical references in the *Rig Veda*, and one of these refers to a solar eclipse, which is the subject of this paper.

## 2 'ATRI'S ECLIPSE' IN THE *RIG VEDA*

In the *Rig Veda* (V, 40, 5–9) there is a mention of a total solar eclipse. We quote the verse from Jemison and Brereton (2014), which states:

- When, O Sun, Svarbhānu Āsura pierced you with darkness like a befuddled man not knowing the territory the living beings perceive. (5)
- Then, O Indra, when you smashed down from heaven the circling magic spells of Svarbhānu, Atri with the fourth formulation found the sun, hidden by darkness because of (an act) contrary to the commandment. (6)
- [The Sun:] "O Atri, let him not, deceived by jealousy and fear, swallow me, who am one of yours. You are an ally whose bounty is real; do you and King Varuṇa help me here." (7)
- The possessor of the sacred formulation, having yoked the pressing stones, serving the gods with plain reverence, doing his utmost, Atri placed the eye of the sun in heaven. He hid away the magic spells of Svarbhānu. (8)
- Which sun Svarbhānu Āsura pierced with darkness, that one the Atris found, for no others were able. (9)

Here is another translation of the verse, this time by Griffith (1896):

- O Surya, when the Asura's descendant Svarbhanu, pierced thee through and through with darkness, All creatures looked like one who is bewildered, who knoweth

- not the place where he is standing. (5)
- What time thou smotest down Svarbhanu's magic that spread itself beneath the sky, O Indra, By his fourth sacred prayer Atri discovered Surya concealed in the gloom that stayed his function. (6)
- Let not the oppressor with this dread, through anger swallow me up, for I am thine, O Atri. Mitra art thou, the sender of true blessings: thou and King Varuna be both my helpers. (7)
- The Brahman Atri, as he set the press stones, serving the Gods with praise and adoration, Established in the heaven the eye of Surya, and caused Svarbhanu's magic arts to vanish. (8)
- The Atris found the Sun again, him whom Svarbhanu of the brood Of Asuras had pierced with gloom. This none besides had the power to do. (9)

Note that neither of the two translations suggests the myth of *Rahu* and *Ketu*, the more common myth of eclipses in India (cf. Vahia and Halkare, 2017). This reinforces the suggestion that this verse is significantly older than the more common story in the *Samudra Manthana*, which is a major episode in Hinduism. It also refers to Indra and not Vishnu, further suggesting that the myth is older than that of *Rahu* and *Ketu*, which is believed to be a Puranic story (e.g. see Dimmitt, 2012).

The reference to Indra as the possessor of the Bull and Vrtra slayer is instructive in defining the astronomical reference-frame. In the *Rig Veda*, the description of the Canis (Major and Minor) and Orion regions defines the idea of Indra (Sirius) killing the bull in Taurus, suggesting the end of the year of Pitras (the journey of the Sun from the Autumnal Equinox to the Spring Equinox) and beginning of a New Year symbolised by Indra making it easy for the Sun to come to the world of Devas (from the Spring to the Autumnal Equinox). This is discussed in detail by Tilak (1893). We discuss Tilak's interpretation below.

## 2.1 Tilak's Interpretation

Tilak (1893: 166–167) gives more details:

The fortieth hymn in the fifth Mandala of the Rigveda is still more important in this connection. It shows that an eclipse of the sun was then first observed with any pretensions to accuracy by the sage Atri … This observation of the solar eclipse is noticed in the Sankhyayana (24, 8) and also in the Tandya Brahmana (iv. 5. 2; 6. 14) in the former of which it is said to have occurred three days previous to the Vishuvan (the Autumnal Equinox).

Regarding the mention that the eclipse had occurred three days prior to the Autumnal Equinox, we quote the original verse in Kaushitiki or Sankhyayana Brahmana. 24.3:

स्वर्भानुर् ह्वा आसुर आदित्यम् तमसा अविध्यत्।
तस्य अत्रयस्तमों अपिजिघांसन्तऐतम्सप्तदच
स्तोमम् त्र्यहम् पुरस्ताद् विशुवत उपायन्।
त ऐतम् एव त्र्यहम् उपरिष्टाद् विशुवत उपायन्।

Keith (1981) gives a translation of the passage. He says that Kaushitaki Upanishad, (Chapter 24, Brahman, i.e.Mantra 3) states that:

Swarbhanu, an Asura pierced with darkness the sun; Atri were fain to smite away its darkness;
they performed before Visuvanta, this set of three days, with Saptadaca Stoma.
They smote away the darkness in front of it; that settled behind;
they performed the same three-day (rite) after Visuvat; they smote way the darkness behind it.

## 3 PARAMETERS OF THE ECLIPSE

Based on this we now have, in the *Rig Veda* (V, 40 verses 5 to 9), a description of the eclipse. Tilak (1893) further specifies the date of the eclipse as being three days before the Autumnal Equinox. We therefore have four details of the eclipse:

1. It occurred when the Vernal Equinox was in Orion;
2. It was a total solar eclipse;
3. It occurred three days prior to the Autumnal Equinox; and
4. It was total wherever the Rig Vedic people were living at that time.

We now look at each of these parameters in detail.

### 3.1 Vernal Equinox in Orion

The Vernal Equinox drifts in the sky by about 83.81 arcmin per century or 50.29 arcsec per year, and hence over a period of about 25,800 years the axis returns to the same point (the rate changes with time, so the precise period cannot be given). Consequently, the astronomical background of the four cardinal points, the equinoxes and the solstices changes with time.

Currently the Vernal Equinox is in Pisces, but around 4000 BC ± 200 it would have been in Orion (Baity, 1973). The association of the Sun or Moon at that time was referred to asterisms and not to the boundaries of a region. Hence the point of the Vernal Equinox had to be within the asterism of Taurus, thus allowing us to define the time period precisely.

### 3.2 Description of Totality

The 5[th] verse quoted above clearly states that when the Asura's descendant Svarbhanu pierced thee through and through with darkness, all creatures looked like one who is bewildered, who knew not the place where he is standing. This clearly describes a total solar eclipse, for a partial eclipse would not have bewildered mankind in this way, as has been confirmed by several interpretations (e.g. see Donier, 1984: 187).

### 3.3 The Autumnal Equinox and the Date of the Eclipse

In this paper, we use dates from the Julian Calendar although they were not used in the era of the *Rig Veda*. It should be noted that since the mean length of one year in the Julian Calendar is 365.25 days, which is longer than the tropical year of 365.2422 days, the dates of the Autumnal Equinox in around 4000 BC were 20 October or so. The precise dates of the Autumnal Equinoxes were calculated as the instants when the apparent ecliptic longitude of the Sun was 180º.

### 3.4 The Location of the Observations

The Rig Vedic were a nomadic ethnic group that lived in Central Asia (Dalal, 2014). In a later book, Tilak (1903) presents evidence that they originally lived in far northern Central Asia, close to the Arctic Circle. This tallies with documentation in the *Rig Veda* (for example RV VII, 76, 3), which shows that the Rig Vedic people experienced several days where only the dawn came and went without the Sun being visible. This would occur amongst people living in the northern shadows of the mountains in high latitudes above about 50° N (ibid.).

On the basis of archaeological, linguistic and genetic data, it is generally believed that the Rig Vedic people came from Central Asia to India around 2000 BC during the decay phase of the Harappan Civilisation (c.f. Joseph, 2018), entering the Subcontinent via the Khyber Pass (Spengler et al., 2014). However, at the time of the eclipse, the Vedic people were residing in Central Asia and hence the eclipse would have been observed from Central Asia. Recent genetic data on migrations from Central Asia and West Asia into the Indian Subcontinent (Narasimhan et al., 2018; 2019) reinforce these suggestions. We, therefore, looked for eclipses that would have been visible from Central Asia, not from the Indian Subcontinent.

### 4 SIMULATING THE ECLIPSE

Based on the above discussions, we assume that the eclipse was seen between 4200 and 3800 BC, was total in Central Asia, and that it occurred three days prior to the Autumnal Equinox. We assume that people at that time could determine the day of the Autumnal Equinox with the accuracy of one day by measuring the Sun's altitude at noon. We use the simulation techniques of Sôma and Tanikawa (2015; 2016) to determine possible eclipses that fitted these criteria. For the eclipse calculations and determination of dates of the Autumnal Equinox, we use JPL's planetary ephemerides DE431 (Folkner et al., 2014) and formulae for the Earth's precession by Simon et al. (1994).

For the eclipse calculations, we need the value of the Earth's rotational clock error $\Delta T$ = TT − UT, where TT is Terrestrial Time and UT is Universal Time (Stephenson, 1997). For the period concerned, we had no idea about the value of $\Delta T$, so we searched all total eclipses which occurred 3 ± 1 days prior to the Autumnal Equinox and we adjusted the value of $\Delta T$ so that the totality was seen in Central Asia. It should be noted that, if the $\Delta T$ value is increased, the area where the eclipse can be seen moves eastward at the rate of 15° × 1.0027379 in longitude per hour in $\Delta T$, where 1.0027379 is the ratio of sidereal time to mean solar time. We only found two possible candidates—their paths of totality are shown in Figures 1 and 2.

The parameters of these eclipses are:

(1) 22 October 4202 BC, when the Autumnal Equinox was at 11:12 UT on 24 October, and assuming $\Delta T$ = 120,000s; and
(2) 9 October 3811 BC, when the Autumnal Equinox was at 09:42 UT on 21 October, and assuming $\Delta T$ = 130,000s.

Both of these eclipses would have been seen as total or nearly total by people living around the Caspian Sea or in Central Asia, whose descendants would eventually migrate to India.

The $\Delta T$ values shown above were obtained from the condition that the total eclipses were seen in Central Asia, but since we do not know the exact observing sites, they should have errors of about ±5,000s. Stephenson et al. (2016) show that the following quadratic equation of time approximates the values of $\Delta T$ since 720 BC:

$\Delta T = \{-320 + 32.5[(\text{year} - 1825)/100]^2\}$s.   (1)

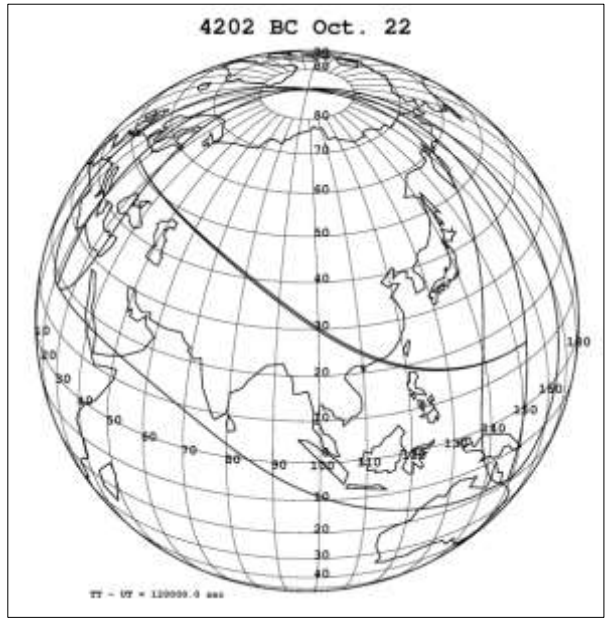

Figure 1: Eclipse path of 4202 BC.

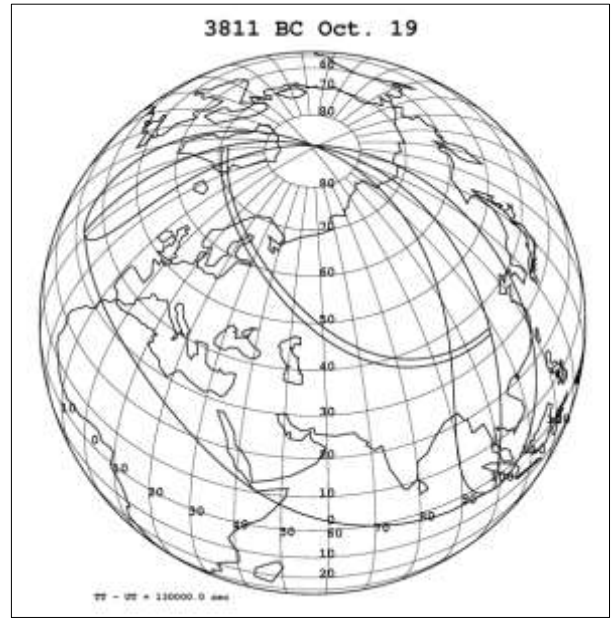

Figure 2: Eclipse path of 3811 BC.

It should be noted that there is no guarantee that this formula can also be applied to eclipses dating to around 4000 BC. But if we do use this formula for the epoch of 'Atri's eclipse', the $\Delta T$ values for 4202 BC and 3811 BC become 118,000s and 103,000s, respectively. So, if Atri's eclipse was the eclipse in 4202 BC, we see that the above quadratic formula of $\Delta T$ gives a good approximation for the value in 4202 BC, but if the eclipse was in 3811 BC the quadratic formula gives a value that is too small by about 27,000s.

## 5  DISCUSSION

The conditions for 'Atri's eclipse' have been given by Tilak, who quoted from the *Rig Veda* and other literature. He concludes that this eclipse occurred when the Vernal Equinox was in Orion, indicating that it dates to 4000 ± 200 yrs. Tilak then goes on to state that it was observed as a total solar eclipse and occurred three days before the Autumnal Equinox. With this specific information, we looked for possible eclipses and found only two that satisfy these three conditions of Vernal Equinox, the Autumnal Equinox and totality visibility in Central Asia.

We then reviewed the option of choosing Central Asia rather than India. The years 4202 and 3811 BC were before the emergence of the Harappan Civilisation in the Indian Subcontinent, and this was a time when the Central Asian region was active (see Narasimhan et al., 2018; 2019). Tilak (1893) mentions several astronomical records in the *Rig Veda* where the observations were made in regions where the Sun only appeared in the sky around the time of the Vernal Equinox, suggesting habitation even north of Central Asia. Lazaridis et al. (2016) point out that farming was carried out in this region as early as 7000 BC, suggesting that there was a significant settled population in the Central Asian region.

Ancient literature (e.g., see Mahajani et al., 2007) is known to include memories of events such as the Rohini Shakat Bheda last occurred more than 7000 years ago, which supports the suggestion that 'Atri's eclipse' mentioned in the *Rig Veda* refers to a total solar eclipse that was observed somewhere in Central Asia in either 4202 BC or 3811 BC.

## 6  CONCLUDING REMARKS

We propose that the eclipse recorded in the *Rig Veda* refers to observations made of an eclipse around 4000 BC. By analyzing the description, we propose that the eclipse was the one that occurred in 4202 BC or else in 3811 BC. We propose that it was observed in Central Asia. To our knowledge, this is one of the oldest known references to a specific total solar eclipse mentioned in the historical literature.

## 7  ACKNOWLEDGEMENTS

The first author of this paper (MNV) wants to acknowledge the support of an old and dear friend, Professor Wayne Orchiston. Over the years he has helped me with my studies of Indian archaeoastronomy and ethnoastronomy, and I would like to dedicate this paper to him.

MNV also would like to acknowledge the valuable discussions with Dr Devadutta Pattanaik and all the references provided by Dr A. Jamkhedkar and Dr Pramod Pandey, Vice Chancellor of Deccan College. The other author (MS) carried out the eclipse calculations reported here using the Multi-Wavelength Data Analysis System operated by the Astronomy Data Center at the National Astronomical Observatory of Japan.

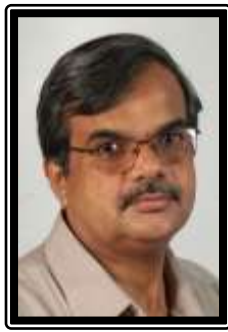

**Professor Mayank Vahia** was until recently a scientist at the Tata Institute of Fundamental Research, Mumbai, and is now Dean of Mathematical Sciences at the Narsee Monjee Institute of Management Studies in Mumbai.

He completed his PhD at the University of Mumbai in 1984 and began his career at the Tata Institute of Fundamental Research with an interest in cosmic rays. He was involved in an experiment that was flown on NASA's Space Shuttle Space Lab 3 mission in 1986. After that he widened his interests and worked on high energy (X-ray and Gamma Ray) telescopes that were flown on Russian and Indian satellites.

For the past fifteen years or so he has been interested in the origin of astronomy in India and has studied the astronomical aspects from early rock art, megaliths, coins, architecture, ancient texts and the astronomy of some of Indian's oldest tribes. He has published about 260 papers, around 60 of which are in the history of astronomy and history of science.

He also has published two books: *History of Indian Astronomy: A Handbook* (2016, Indian Institute of Technology and Tata Institute of Fundamental Research, co-edited by K. Ramasubramanian and Aniket Sule), and

*The Growth and Development of Astronomy and Astrophysics in India and the Asia-Pacific Region: ICOA-9, Pune, India, 15-18 November 2016* (2019, Hindustan Book Agency and Springer Singapore, co-edited by Wayne Orchiston and Aniket Sule).

Mayank also spearheaded India's participation in the International Astronomy Olympiad, a programme that he initiated in India and that has guided about 30 students to pursue their studies in science for a career.

**Dr Mitsuru Sôma** was born in Kuroiso (Tochigi Prefecture, Japan) in 1954, and has MSc and PhD degrees in Astronomy from the University of Tokyo. He is currently a VisitingProfessor at the National Astronomical Observatory of Japan. Mitsuru was an Organizing Committee member of IAU Commission 41 (History of Astronomy) during 2009–2015. He is also a member of IAU Divisions A (Fundamental Astronomy), B (Facilities, Technologies and Data Science), C (Education, Outreach and Heritage) and F (Planetary Systems and Bioastronomy). In addition, he is also a Vice President for Grazing Occultation Services of the International Occultation Timing Association.

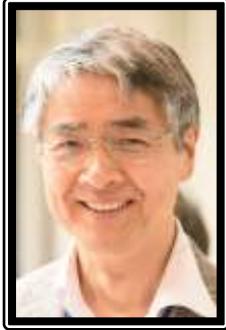

His research interests include linkage of stellar reference frames with dynamical reference frames using observations of lunar occultations and changes in the Earth's rotation during ancient times using ancient records of eclipses and occultations.

Mitsuru has many publications in history of astronomy, include the book *Mapping the Oriental Sky. Proceedings of the Seventh International Conference on Oriental Astronomy (ICOA-7)* (2011, co-edited by Tsuko Nakamura, Wayne Orchiston and Richard Strom).